\documentclass[12pt]{article}
\usepackage{techart}

\renewcommand\vec[1]{\boldsymbol{#1}}
\renewcommand\mat[1]{\boldsymbol{#1}}
\newcommand\transpose[1]{\ensuremath{{#1}^T}}

\newcommand\x{x}
\newcommand\y{y}
\newcommand\Y{Y}

\newcommand\vx{\vec{\x}}
\newcommand\vy{\vec{\y}}

\newcommand\mX{\mat{X}}

\newcommand\vz{\vec{z}}
\newcommand\mH{\mat{H}}

\newcommand\fy{\vec{\tilde{\y}}}

\newcommand\trend{\beta}
\newcommand\nugget{\tau}
\newcommand\scale{\lambda} 
\newcommand\smoothness{\kappa}
\newcommand\sd{\eta}
\newcommand\anis{\mat{A}}
\renewcommand\angle{\alpha}

\newcommand\mR{\mat{R}}
\newcommand\mSigma{\Sigma}

\newcommand\gauss{\phi}

\newcommand\assign{\ensuremath{\Leftarrow}}

\begin{document}
\title{%
Iterative Posterior Inference for Bayesian Kriging}
\author{Zepu Zhang}
\date{October 13, 2011}
\maketitle

\barefoot{zepu.zhang@gmail.com\\
Department of Mathematics and Statistics, University of Alaska,
Fairbanks, Alaska, USA\\[5pt]
Published in \texttt{Stochastic Environmental Research and Risk
Assessment}, 2011. doi:10.1007/s00477-011-0544-y}

\vspace*{-10mm}

\begin{abstract}
We propose a method for estimating the posterior distribution
of a standard geostatistical model.
After choosing the model formulation and specifying a prior,
we use normal mixture densities to approximate the posterior
distribution.
The approximation is improved iteratively.
Some difficulties in estimating the normal mixture densities,
including determining tuning parameters concerning bandwidth and
localization, are addressed.
The method is applicable to other model formulations as long as all the
parameters, or transforms thereof, are defined on the whole real line,
$(-\infty, \infty)$.
Ad hoc treatments in the posterior inference such as
imposing bounds on an unbounded parameter or
discretizing a continuous parameter are avoided.
The method is illustrated by two examples,
one using digital elevation data and the other using historical soil
moisture data.
The examples in particular examine
convergence of the approximate posterior distributions in the
iterations.

\textbf{Key Words:}
geostatistics; importance sampling; normal mixture;
kernel density estimation; anisotropy; change of support
\end{abstract}

\section{Introduction}

In the geostatistical literature,
likelihood-based methods for parameter estimation
make explicit assumptions on the
statistical distributions of the spatial variable in question,
thus contrast with some traditional methods
such as ones based on curve-fitting for an empirical variogram model
\citep{Kovitz:2004:SSC, Emery:2007:RFS}.
The distributional assumptions are manifested, for example,
in ``model-based geostatistics''
\citep{Diggle:2007:MBG}
and ``trans-Gaussian'' models
\citep{Christensen:2001:APV}.
First employed apparently in the early 1980s,
likelihood-based methods have by now established themselves
as a standard approach \citep[see][for a review]{Zimmerman:2010:LBM}.

Given the distributional assumptions,
one may obtain maximum likelihood (ML) estimates of the parameters,
such as parameters in the spatial covariance function.
However,
estimating an ``optimal'' value for the parameters of the covariance function
via ML methods is flawed because,
according to \citet{Warnes:1987:PLE},
the profile likelihood can be multimodal
and is often nearly flat in the neighborhood of its mode.
These problems suggest that
ML estimates may be difficult to find and
such estimates, if found, may not be truly ``representative'' or ``optimal''.
\citet{Handcock:1993:BAK} further emphasize that
``plug-in'' measures of prediction uncertainty based on ML parameter
estimates tend to be over optimistic.
They argue that a Bayesian approach mitigates this problem.
Indeed,
parallel to the growth of Bayesian statistics in general,
Bayesian geostatistics (or Bayesian kriging) has enjoyed steady growth
since the 1980s
\citep{%
Kitanidis:1986:PUE,
Handcock:1993:BAK,
deOliveira:1997:BPT,
Ecker:1999:BMI,
Berger:2001:OBA,
Diggle:2002:BIG,
Banerjee:2004:HMA,
Palacios:2006:NGB,
Cowles:2009:RMP}.

In this study,
we take a typical geostatistical formulation
and present an algorithm
for deriving a numerical
representation of the posterior distribution of the parameters.
The parameter vector, $\theta$,
consists of standard elements such as
trend coefficients $\trend$,
variance $\sd^2$,
scale $\scale$,
smoothness $\kappa$, etc.
It also accommodates geometric anisotropy.
The prior specification combines
standard non-informative priors (for $\trend$ and $\sd^2$)
and informative priors (for $\scale$ and $\kappa$).
The algorithm uses a normal mixture to approximate the posterior density;
the approximation is updated iteratively to approach the true
posterior distribution.

The proposed algorithm avoids two empirical treatments
that have been used in the literature,
namely, imposing bounds on an unbounded parameter
and discretizing a continuous parameter.
One parameter that has received such treatments
is the scale parameter, $\scale$
\citetext{%
\citealp{deOliveira:1997:BPT};
\citealp[sec.~7.2]{Diggle:2007:MBG};
\citealp{Cowles:2009:RMP}}.
These treatments have conceptual as well as practical complications.
Conceptually,
one needs to take great care to defend one's subjectively chosen bounds
for an unbounded parameter by showing that the posterior inference
is not sensitive to these bounds
\citep{Berger:2001:OBA}.
Practically,
discretizing a continuous parameter renders the computational cost
directly proportional to the resolution of the discretization.
If discretization and artificial bounds are applied simultaneously
on one parameter, say the scale $\scale$,
the desires to use wide bounds and dense discretization
exert great burden on computation.
Furthermore,
the discretization approach is not scalable,
in the sense that the computational cost grows exponentially
as the number of parameters being discretized increases.

The central step of the algorithm,
multivariate kernel density estimation,
is a standard task with unsolved difficulties
\citep{Silverman:1986:DES, Scott:1992:MDE, Wand:1995:KS}.
Our algorithm makes particular efforts to determine two tuning
parameters,
one concerning localization and the other concerning bandwidth,
by a likelihood criterion.
Another technical difficulty arises from high skewness of
importance weights, especially in early iterations.
This problem is alleviated by a ``flattening'' transform.

This study intends to provide a generic, or routine,
procedure for the posterior inference of Bayesian kriging parameters.
As already mentioned,
the procedure specifies a prior with limited user intervention,
improves an initial, rough approximation iteratively,
and avoids some ad hoc treatments that have been used in the literature.
Moreover,
the procedure is readily applicable to other model formulations
so long as all the model parameters are defined on
$(-\infty, \infty)$ or can be transformed to be so.
In this sense,
application of the algorithm extends beyond geostatistics.

The article is organized as follows.
The geostatistical parameterization for a spatial variable
is described in Section~\ref{sec:parameterization},
with details on the Mat{\'e}rn correlation function,
geometric anisotropy,
the likelihood function,
and specification of the prior.
The technical core of this article,
the iterative algorithm for posterior inference,
is the subject of Section~\ref{sec:algorithm}.
The proposed method is capable of dealing with a type of
``change-of-support'' problems; this is briefly discussed in
Section~\ref{sec:change-of-support}.
In Section~\ref{sec:examples},
we apply the geostatistical model and the algorithm
in two examples, one using synthetic data and the other using historical
data.
Section~\ref{sec:concl} concludes the article with a summary.

\section{Parameterization}
\label{sec:parameterization}

Let $\Y(\x)$ be a spatial random variable,
where
$\x \in \Omega \subset \mathcal{R}^d$ is location
in $d$-dimensional space (hence $\x$ is a $d$-vector, $d=1,2,3$).
We adopt the following mixed-effects model for $\Y(\x)$:
\begin{equation}\label{eq:Y-mixed-model}
\Y(\x)
= \transpose{\mu(\x)}\trend
    + \sqrt{1 - \nugget} \,\sd \,\varrho(\x)
    + \sqrt{\nugget}\sd \,\epsilon(\x)
,
\end{equation}
where
$\mu(\x)$ is a $p$-vector of deterministic covariates
(e.g.\@ polynomials of the spatial coordinates);
$\trend$ is a $p$-vector of trend coefficients;
$\varrho(\x)$ is a zero-mean, unit-variance, stationary Gaussian
process;
$\epsilon(\x)$ is an iid standard normal white-noise process;
$\sd > 0$ is the standard deviation of $\Y(\x)$;
and
$0 \le \nugget < 1$ is a nugget parameter.
The process $\varrho(\x)$ is characterized by a
correlation function parameterized by a $q$-vector $\phi$.
In addition, we assume $\varrho(\x)$ and $\epsilon(\x)$ are independent
of each other.

We denote the full parameter vector by
$\theta = (\trend, \nugget, \sd^2, \phi)$.
Modeling efforts are typically concentrated on inferencing and
interpreting
the nugget parameter $\nugget$,
the variance $\sd^2$,
and
the correlation parameter(s) $\phi$.
The content of $\phi$ depends on the specific correlation
function employed;
our choice here is the Mat\'ern correlation function,
to be described below.
In addition,
the formulation above allows for geometric anisotropy,
parameters of which are also contained in $\phi$.

\subsection{Correlation function}

We assume spatial stationarity for $\Y$,
hence the correlation between $\Y(\x_1)$ and $\Y(\x_2)$
is a function of
scaled distance, denoted by
$\ell(\x_1,\x_2) = |\x_1 - \x_2| / \scale$,
where $\scale$ is the ``scale'' parameter.

The marginal distribution of $\Y(\x)$
in the formulation~(\ref{eq:Y-mixed-model})
is
$N\bigl(\transpose{\mu(x)}\!\trend,\, \sd^2\bigr)$.
Of the total variance $\sd^2$,
$\nugget \sd^2$ is contributed by the white noise component
$\sqrt{\nugget}\sd\, \epsilon(\x)$,
whereas the spatially correlated component $\sqrt{1-\nugget}\sd
\varrho(\x)$ contributes a variance of $(1-\nugget)\sd^2$.
In other words,
the so-called ``nugget effect'' accounts for a fraction $\nugget$
of the total variance.
The covariance between
$\Y(\x_1)$ and $\Y(\x_2)$ is
\begin{equation}\label{eq:cov}
\begin{split}
\cov\bigl(\Y(\x_1), \Y(\x_2)\bigr)
&= \sd^2 \corr\bigl(\Y(\x_1), \Y(\x_2)\bigr)
\\
&= \sd^2 \Bigl(
    (1 - \nugget) \rho\bigl(\ell(\x_1,\x_2)\bigr)
    + \nugget I(\x_1 = \x_2)
    \Bigr)
,
\end{split}
\end{equation}
where $I$ is the identity function,
assuming value 1 if $\x_1$ coincides with $\x_2$ and 0 otherwise,
and $\rho$ is taken to be the Mat\'ern correlation function:
\begin{equation}\label{eq:matern-corr}
\rho(\ell; \smoothness)
= \frac{1}{2^{\smoothness - 1} \Gamma(\smoothness)}
    \,
    \ell^{\smoothness}
    \,
    \mathcal{K}_{\smoothness}(\ell)
,\quad
\smoothness > 0
,
\end{equation}
where $\Gamma$ is the gamma function,
$\mathcal{K}_{\smoothness}$ is the modified Bessel function
of the third kind of order $\smoothness$
\citep[secs~9.6 and 10.2]{Abramowitz:1965:HMF};
and
$\smoothness$
is the smoothness parameter
\citetext{%
\citealp[p.~31]{Stein:1999:ISD};
\citealp[p.~51]{Diggle:2007:MBG}}.

In summary,
in a 1-D model or 2- and 3-D isotropic models,
the correlation parameter $\phi$ contains two elements,
$\scale$ and $\smoothness$.
In an anisotropic model,
$\phi$ contains additional elements that characterize anisotropy, as we
discuss next.

\subsection{Geometric anisotropy}

When $d > 1$, our formulation allows for geometric anisotropy.
\citet{Zimmerman:1993:ALA} distinguishes three forms of geometric
anisotropy, including anisotropy in sill,
in range (the ``scale'' parameter here), and in nugget, respectively.
Range anisotropy is the most commonly discussed and is also the
anisotropy considered here.
When such anisotropy is present,
the scaled distance, $\ell$, is calculated in a transformed coordinate
system, which is obtained by
rotating the ``natural'' axes to the major and minor
directions of anisotropy,
and using different scales ($\scale$'s) along different axes.
Specifically,
in 2-D we need one angle ($\angle$) and two scales
($\scale_1$, $\scale_2$) to describe such anisotropy,
whereas in 3-D we need three angles
($\angle_1$, $\angle_2$, $\angle_3$) and three scales
($\scale_1$, $\scale_2$, $\scale_3$).
The rotational angles determine a transformation matrix, $\anis$
\citep[see, \eg][p.~62]{Wackernagel:2003:MG}.
With the matrix $\anis$ and directional scales $\scale_1,\dotsc,\scale_d$, define
\[
\mat{B}(\scale, \anis)
= \anis
  [\diag(\scale_1^{-2},\dotsc,\scale_d^{-2})]
  \transpose{\anis}
,
\]
where
$\diag(\scale_1^{-2},\dotsc,\scale_d^{-2})$ denotes the diagonal matrix
with $\scale_1^{-2},\dotsc,\scale_d^{-2}$ on the main diagonal.
The scaled distance is then defined by
\begin{equation}\label{eq:ell}
\ell(\x_1, \x_2)
= \sqrt{
    \transpose{[\x_1 - \x_2]}
    \mat{B} \,
    [\x_1 - \x_2]}
.
\end{equation}
Note that the $\x_1 - \x_2$ is a column vector of length $d$.
This $\ell$ is used in~(\ref{eq:matern-corr}) to calculate
$\rho$.

In a study of geometric anisotropy in 2-D,
\citet{Ecker:1999:BMI} treat the matrix $\mat{B}$ as parameter and derive
$\anis$ and $\scale$ from $\mat{B}$.
They discuss, in a Bayesian context,
how to specify a prior for
$\mat{B}$ using the Wishart distribution.
The parameterization we adopt goes in the opposite direction.
With our parameterization,
priors for the angle(s) $\angle$
and the scales $\scale$ are specified.
This parameterization has some advantages in terms of
interpretation and intuition.
These two parameterizations contain the same number of unknowns.
\citet{Hristopulos:2002:NAC}
studies more general forms of anisotropy that are not considered here.

In summary,
in a 2-D anisotropic model, the correlation parameter
$\phi$ consists of elements $\smoothness$, $\angle$,
and $\scale_1$, $\scale_2$;
in a 3-D anisotropic model, $\phi$ consists of
$\smoothness$, $\angle_1$, $\angle_2$, $\angle_3$,
and $\scale_1$, $\scale_2$, $\scale_3$.

\subsection{Likelihood}

The parameter vector is denoted by
$\theta = (\trend, \nugget, \sd^2, \phi)$.
The content of the correlation parameter $\phi$
depends on the spatial dimension $d$ and whether anisotropy is
considered, as discussed above.
Suppose we have measurements of $\Y$,
denoted by $\vy$,
at $n$ locations $\vx$.
The likelihood function of $\theta$ with respect to $\vy$ is
equal to
\begin{equation}\label{eq:likelihood}
p(\vy \given \theta)
= (2\pi \sd^2)^{-n/2} \,
   |\mR|^{-1/2}
   \exp\Bigl( -\frac{1}{2\sd^2}
       \transpose{\bigl(\vy - \mX \trend\bigr)}
       \negthinspace
       \mR^{-1}
       \bigl(\vy - \mX\trend\bigr)
       \Bigr)
,
\end{equation}
where
$\mX$ is the ``design matrix'' of covariates corresponding to $\vx$,
each row being $\mu(\x)$ for a single location $\x$;
$\mR$ is the correlation matrix between the locations $\vx$,
calculated using the relations
(\ref{eq:cov}) and~(\ref{eq:matern-corr}).

In applications,
the parameter vector $\theta$ could be simplified depending on the actual
situation of the spatial variable and emphasis of the investigation.
For example, one might choose to fix the smoothness $\smoothness$
at a certain value, say 0.5 or 1.5.
(Then, $\smoothness$ would not appear in $\theta$.)
For another example,
one might decide, based on background knowledge,
not to consider anisotropy,
hence $\scale$ would be a scalar,
and $\phi$ would not contain $\angle$.

When anisotropy is considered,
the coordinate rotation affects location-aware calculations
including the trend function (via $\mu(\x)$ and $\trend$) and
the correlation.
We choose to define the trend function in terms of the original coordinates.
The angles ($\angle$'s) define the rotations;
the scales ($\scale$'s) are applied along the axes after the rotation.
The parameters $\smoothness$, $\nugget$, and $\sd^2$ do not have directional
components because we consider range anisotropy only.

\subsection{Specification of prior}
\label{sec:prior}

We specify a prior with independent components:
\begin{equation}\label{eq:prior}
\pi(\trend, \sd^2, \nugget, \scale, \smoothness, \angle)
=
  \bigl(\sd^2\bigr)^{-1}
  \operatorname{beta}(\nugget; \cdot)\,
  \operatorname{gamma}(\scale; \cdot) \,
  \operatorname{gamma}(\smoothness; \cdot)\,
  \operatorname{unif}(\angle; 0, \pi/2)
.
\end{equation}
(Remember that $\angle$ appears in anisotropic models only.
In anisotropic models, $\scale$ contains, and $\angle$ may contain,
more than one elements.)
This specifies a flat prior on $(-\infty, \infty)$ for the trend coefficients
$\trend$ and a conventional noninformative prior for the variance
$\sd^2$.
It is sensible to use a diffuse prior for $\scale$,
whereas subjective information (or preference) about
$\smoothness$ and $\nugget$ may be injected into their priors.
Relevant discussions can be found
in
\citet{Berger:2001:OBA},
\citet[sec.~5.1.1]{Banerjee:2004:HMA},
\citet[p.~50]{Gelman:2004:BDA},
and
\citet{Gelman:2006:PDV}.
Additional parameters need to be chosen
for the beta and gamma distributions.
Some details are listed below.
However, bear in mind that these particularities are empirical and
subject to adjustment.

$\operatorname{beta}(\nugget;\cdot)$:
this is taken to be
$\operatorname{beta}(\nugget; 1, 5)
= (1 - \nugget)^4 I(0 \le \nugget \le 1)$.
This prior places more weight on small nugget values.

$\operatorname{gamma}(\scale; \cdot)$:
this is taken to be
$\operatorname{gamma}(\scale; 1, L/(2\log 2))$,
where $L$ is the size of the model domain.
This is actually the exponential distribution with median
$L/2$.

$\operatorname{gamma}(\smoothness;\cdot)$:
parameters of this gamma distribution are chosen such that
its mode is 1.5 and its variance is 4.0.

\section{Inference of the posterior}
\label{sec:algorithm}

We estimate the posterior distribution of the parameter vector,
$\theta$, by normal mixtures in an iterative procedure.
The versatility of normal mixtures in approximating complex densities is
documented by \citet{Marron:1992:EMI}.
The normal kernel
implies that all components of $\theta$ must be defined
on $(-\infty, \infty)$.
This requires that the support of each component of $\theta$
is one of
$(-\infty, \infty)$, $(c, \infty)$, $(-\infty, c)$,
and $(c_1, c_2)$, where $c$, $c_1$, and $c_2$ are constants.
Parameters defined on half-bounded intervals
(such as $\sd^2$, $\scale$, and $\smoothness$)
may be log-transformed.
Parameters defined on bounded intervals
(such as $\nugget$ and $\angle$)
may be logit-transformed.
In fact,
the algorithm below is applicable to any model formulation
as long as all the parameters are defined on $(-\infty, \infty)$,
either directly or after transformation.

The prior given by~(\ref{eq:prior}) is for the parameters
on their natural scale.
The prior for the transformed parameters
are determined by (\ref{eq:prior}) and the transformations.
To avoid clutter in notation, we shall still use $\theta$
to denote the parameter vector (now transformed)
and refer to (\ref{eq:prior}) for its prior,
although in reality the prior of the transformed parameter
is a modified form of (\ref{eq:prior}).
The algorithm actually derives posterior distribution of these
\emph{transformed} parameters.
Distribution of the parameters on their natural scale
can be studied based on back-transformed samples
from the derived posterior distribution.

\subsection{Algorithm}
\label{sec:algor}

We begin with an initial approximation to the posterior,
denoted by $f^{(0)}(\theta)$,
which is taken to be a sufficiently diffuse
multivariate normal distribution.
In the $k$th iteration, the current approximation $f^{(k-1)}$
is updated to $f^{(k)}$ in three steps as follows.

\begin{enumerate}
\item
Take a random sample,
$\{\theta_1,\dotsc,\theta_n\}$, from $f^{(k-1)}$.

\item
For $i=1,\dotsc,n$,
compute the non-normalized posterior density
$s_i = \pi(\theta_i)\, p(\vy \given \theta_i)$
and the proposal density
$t_i = f^{(k-1)}(\theta_i)$;
let
$w_i = \frac{s_i / t_i}{\sum_{j=1}^n s_j/t_j}$
be the importance weight of $\theta_i$.

\item
Update the approximate posterior from $f^{(k-1)}$ to
\begin{equation}\label{eq:mixture-normal}
f^{(k)}(\theta)
\approx
\sum_{i=1}^n w_i\, \gauss(\theta;\, \theta_i, V_i).
\end{equation}
This is a mixture of $n$ normal densities (denoted by $\gauss$),
each with mean $\theta_i$ and covariance matrix $V_i$.
Computation of $V_i$ is detailed in
Section~\ref{sec:localization}.
\end{enumerate}

This algorithm does not require one to be able to draw a random sample
from the prior of $\theta$.
Instead, a convenient initial approximation,
$f^{(0)}$,
starts the procedure.
One only needs to be able to \emph{calculate} the prior density for any
particular value of $\theta$.
This provides great flexibilities in choosing
the prior $\pi(\theta)$ and the initial approximation $f^{(0)}$.
Sampling from a normal mixture distribution is easy.

Note that $f^{(k)}$ is a normal mixture, just like $f^{(k-1)}$,
and is ready to be updated in the next iteration.
Alternatively,
one may terminate the iteration
according to certain empirical criterion,
and take $f^{(k)}$ as the final approximate posterior distribution
of the parameter $\theta$.

Some properties of $f^{(k)}$
may be examined semi-analytically.
More often, one is more interested in the unknown field $\Y$ or a function
thereof than in the parameter $\theta$ itself.
Properties of $\Y$ or a function thereof
may be investigated via sampling $\theta$ from $f^{(k)}(\theta)$
and simulating (i.e.\@ sampling) $\fy$ (realization of the field)
according to~(\ref{eq:likelihood}).

\subsection{Algorithm detail: computation of $V_i$}
\label{sec:localization}

The step~3 of the algorithm
entails kernel density estimation,
which is a standard but un-settled task
\citep{Silverman:1986:DES, Scott:1992:MDE, Wand:1995:KS}.
The covariance matrix $V_i$ may be expressed as
$h_i \mSigma_i$,
where
$\mSigma_i$ is the empirical weighted covariance matrix
of the sample points ($\theta$'s) in a certain neighborhood of
$\theta_i$,
and $h_i$ is a ``bandwidth'' parameter.
While $\mSigma_i$ specifies the shape of the kernel centered at
$\theta_i$,
the bandwidth $h_i$ further adjusts the spread of this kernel.

Several factors complicate the computation of $V_i$.
First,
the distribution of the importance weights, $w_i$,
can be highly skewed,
especially in early iterations
when the proposal distribution tends to be very different
from the true posterior distribution.
In not-so-rare pathological cases,
a few sample points (or a single sample point) carry a dominant fraction
of the total weight, making the other sample points negligible.
When this happens,
$\mSigma_i$ may contain variance entries that are nearly 0.
To mitigate this problem,
we use a ``flattened'' version of weights in the computation of $\mSigma_i$.
Let
\begin{equation}\label{eq:flatten-by-entropy}
v_i = \frac{w_i ^ \gamma}{\sum_{j=1}^n w_j^\gamma},
\quad\text{where}\quad
\gamma
= -\frac{1}{\log n} \sum_{i=1}^n w_i \log w_i.
\end{equation}
The exponent $\gamma$ is the ``entropy'' of $\{w_i\}$,
a measure of the uniformity of the weights
\citep[see][]{West:1993:APD}.
If $\{w_i\}$ are all equal, then $\gamma = 1$.
At the other extreme,
if one $w_i$ is 1 and all the other weights are 0,
then $\gamma = 0$.
Note that the weights
$\{v_i\}$ are used in calculating the empirical weighted covariance
matrix $\mSigma_i$;
they do not replace the weights
$\{w_i\}$ in~(\ref{eq:mixture-normal}).
As the algorithm proceeds in iterations,
the importance weights $\{w_i\}$ become more uniform,
hence $\gamma$ becomes closer to 1,
and the adjustment to $\{w_i\}$ by the above ``flattening'' becomes minor.

The second complication is in the ``localization'' of $\mSigma_i$,
that is,
we choose to define $\mSigma_i$ as the empirical weighted covariance matrix
of sample points in a ``neighborhood'' of $\theta_i$,
say sample points that take up a fraction $r_i$,
$0 < r_i \le 1$, of the entire sample.
Such localization is important if
the target distribution (i.e.\@ the posterior) is severely multi-modal
\citep{Givens:1996:LAI}.
However, there is no guidance on the determination of the fraction $r_i$.

The third complication is due to the bandwidth $h_i$.
A number of adaptive procedures have been proposed for choosing $h$
\citep{Jones:1990:VKD, Hall:1991:ODB, Sheather:1991:RDB, Terrell:1992:VKD,
Givens:1996:LAI, Sain:2002:MLA}.
However, most of the literature in kernel density estimation
is developed based on a \emph{random} sample,
whereas what we have here is a \emph{weighted} one.
In addition,
localization, as parameterized by $r_i$,
has received less attention in the literature.
Traditional rule-of-thumb choices for the bandwidth parameter
\citep{Jones:1996:BSB}
do not apply directly to a localized algorithm,
because the rules are based on analysis of global estimators.
A localized algorithm encounters other difficulties,
such as edge effects,
that do not arise in a global analysis.

To simplify the matter,
we use a common bandwidth parameter, denoted by $h$ (where $h > 0$),
and a common localization parameter, denoted by $r$ (where $0 < r \le
1$),
for all the mixture components.
Sensible choices
of these two tuning parameters depend on the sample size $n$,
the dimensionality of $\theta$,
and characteristics of the target distribution $\pi(\theta)p(\vy\given
\theta)$.
We determine their values by a maximum likelihood cross-validation criterion:
\begin{equation}\label{eq:ML-kernel}
(r, h)
= \argmax_{r,h}
    J(r, h;\, \theta_1,\dotsc,\theta_n)
,
\end{equation}
where
\begin{equation}\label{eq:obj-kernel}
J(r, h;\, \theta_1,\dotsc,\theta_n)
= \sum_{i=1}^n
    w_i
    \log\biggl(
        \frac{1}{1 - w_i}
        \sum_{j\ne i}
            w_j\,
            \gauss\bigl(\theta_i;\; \theta_j, h \mSigma_j\bigr)
        \biggr)
.
\end{equation}
Note that
$\mSigma_j$ is a function of
$r$, $\{\theta_i\}$, and $\{v_i\}$.
In words,
$J$ is the usual log likelihood of $r$ and $h$ with respect to
the weighted sample $\{\theta_i\}$,
except that the density at $\theta_i$ is calculated by the
mixture density \emph{leaving out} the mixture component centered at
$\theta_i$.
Leaving out the offending mixture component is key.
Otherwise, maximizing $J$ would drive $h$ to be arbitrarily small.
The idea above is quite general,
and may be extended to determine other tuning parameters.

To reduce the computational cost of this optimization,
a combination of discrete search for $r$
and continuous search for $h$
is performed as sketched below.

\begin{enumerate}
\item
Let $J^* \assign -\infty$,
that is, assign value $-\infty$ to the variable $J^*$.

\item
Let $r \assign 1$.

Let
$\Theta_1 \assign \dotsb \assign \Theta_n
\assign \{\theta_1,\dotsc,\theta_n\}$.

Compute the empirical weighted covariance matrix of the entire sample
$\{\theta_1,\dotsc,\theta_n\}$,
using the flattened weights $\{v_i\}$
instead of the original weights $\{w_i\}$.
Denote the result by $\mSigma$,
and let $\mSigma_1 \assign \dotsb \assign \mSigma_n \assign \mSigma$.

\item\label{it:optim-h}
Find $h$ that maximizes~(\ref{eq:obj-kernel}).
Denote the maximizing $h$ by $h_*$
and the achieved maximum $J$ by $J_*$.

(This is a univariate optimization problem.
Since we have values of $\mSigma_1$,..., $\mSigma_n$,
the parameter $r$ does not appear in the calculation
of~(\ref{eq:obj-kernel}).)

If $J_* > J^*$, then
let
$J^* \assign J_*$,
$r^* \assign r$,
$h^* \assign h_*$,
and
$\mSigma_i^* \assign \mSigma_i$
for $i=1,\dotsc,n$.

\item\label{it:halve-r}
If $r$ is below a pre-specified threshold fraction,
say $\frac{1}{8}$,
go to step~\ref{it:conclude}.
Otherwise,
let $r \assign r/2$
and go to step~\ref{it:localize}.

\item\label{it:localize}
For $i=1,\dotsc,n$,
\begin{enumerate}
\item
Within the set $\Theta_i$,
identify the $\lceil r n\rceil$ closest neighbors of
$\theta_i$ in the Mahalonobis sense measured by
the covariance matrix $\mSigma_i$.
Update $\Theta_i$ to be the set that contains
these newly identified neighbors.
\item
Compute the empirical weighted covariance matrix,
$\mSigma_i$, based on the sample $\Theta_i$
and the corresponding relative weights $\{v_i\}$.
\end{enumerate}

Go to step~\ref{it:optim-h}.

\item\label{it:conclude}
Adopt
$r^*$ and $h^*$ as the final values for $r$ and $h$,
respectively,
and let the $V_i$ in~(\ref{eq:mixture-normal}) be
$h^* \mSigma_i^*$ for $i=1,\dotsc,n$.

This concludes the search for optimum values of
$r$ and $h$.
\end{enumerate}

The ideas of normal mixture and iterative updating
are used by \citet{West:1993:APD}.
The procedure in \citet{West:1993:APD}
sets the bandwidth parameter following empirical rules
and does not consider localization.

\section{Provisions for the ``change of support'' problem}
\label{sec:change-of-support}

The data $\vy$ in~(\ref{eq:likelihood}) are the values of $\Y$
at individual locations $\vx$.
This formulation can be easily generalized to use data
that are \emph{linear} functions of $\Y$.
Let the data vector, denoted by $\vz$,
be expressed as
\[
\vz = \mH \vy,
\]
where
$\vy$ is a $n$-vector of $\Y$ at locations 
$\vx$
and
$\mH$ is a $m\times n$ matrix of rank $m$, where $m \le n$.
Correspondingly,
the likelihood~(\ref{eq:likelihood}) is replaced by
\begin{equation}\label{eq:likelihood-linear}
p(\vz \given \theta)
= (2\pi \eta^2)^{-m/2} \,
   |\mH\mR\transpose{\mH}|^{-1/2}
   \exp\Bigl( -\frac{1}{2\eta^2}
       \transpose{\bigl(\vz - \mH \mX \trend\bigr)}
       [\mH\mR\transpose{\mH}]^{-1}
       \bigl(\vz - \mH \mX \trend\bigr)
       \Bigr)
,
\end{equation}
where $\mX$ is the design matrix for the locations $\vx$,
and $\mR$ is the correlation matrix between the locations $\vx$.
The algorithm described in Section~\ref{sec:algorithm}
works for this form of data and likelihood without modification.

In applications,
measurements of $\Y$ often have areal or volume (or, in 1-D, interval)
support rather than point support.
Consider studies in which the model domain is discretized into a numerical grid.
Let us call the numerical grid the ``basic'' spatial unit (or support)
in the model.
It may occur that a data value is some function
(e.g.\@ the average) of $Y$ in more than one basic spatial unit.
Specifically,
we may distinguish 
``point data'', which have the basic support in the model,
and
``linear data'', which have aggregated support and can be expressed as
linear functions (such as simple averages) of point data.
Increasingly in environmental studies,
available data include a mix of
satellite- and ground-based measurements that span a hierarchy of spatial
supports.
This entails the so-called ``change of support'' problems
\citep{Young:2007:LSD},
a typical example being ``downscaling'' problems.
The method described above is able to use data on a variety of supports
as long as they are all \emph{linear} functions of $\Y$ on the basic
support.

Besides being a natural situation due to data resources on disparate
scales,
linear data can also be useful by methodological design.
For example,
\citet{Zhang:2011:AAI} proposes an inverse algorithm in which
a key model device is linear functions of the random field.

\section{Examples}
\label{sec:examples}

We illustrate the algorithm with two examples.
The first example uses a satellite elevation dataset to demonstrate
the algorithm's performance in a 2-D, anisotropic setting.
Because the true field is known in this case,
model performance can be assessed by comparing conditional simulations
with the true field.
The second example uses a historical dataset of soil moisture.
In this realistic setting, the true field is unknown,
and the sampling locations of the measurements are not ideal as far as
interpolation is concerned
(as is a common situation with historical data).
However, the point of the examples is not to interpolate,
but to illustrate how the algorithm
works with available data to approximate the posterior distribution of the
model parameters in an iterative procedure.

\subsection{Example 1}
\label{sec:example1}

\begin{figure}
\centering
\includegraphics[scale=.9]{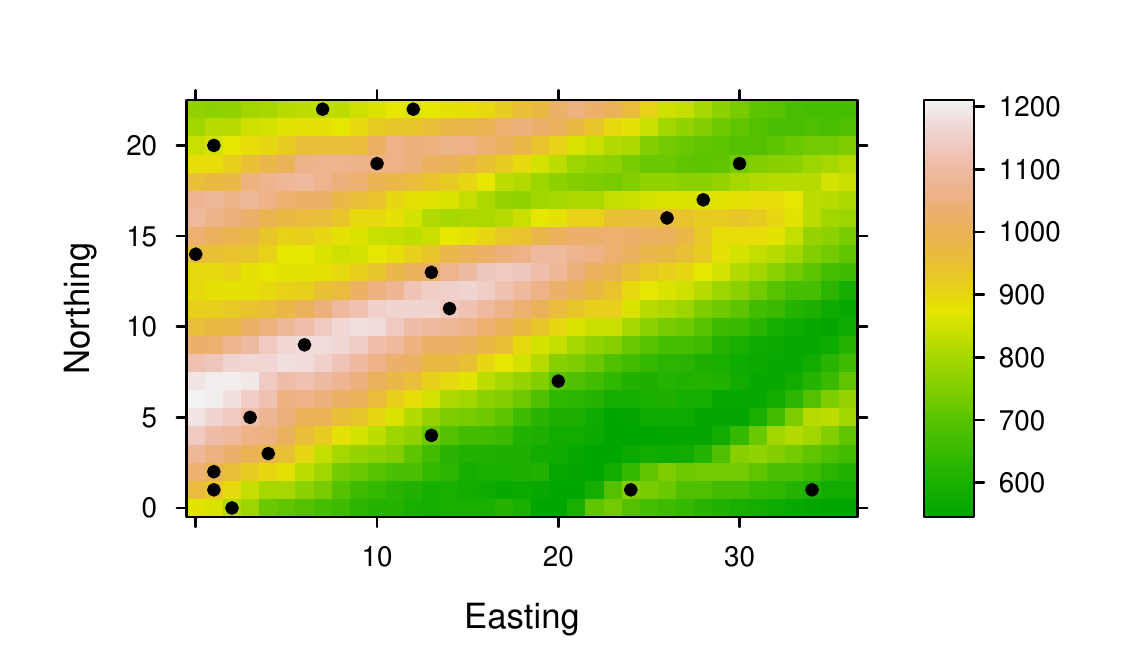}
\caption{Elevation map as the synthetic true field
    for the example in Section~\ref{sec:example1}.
    The 20 dots indicate synthetic measurements.}
\label{fig:appal-datamap}
\end{figure}

We extracted satellite tomography data from the National Elevation Dataset
(NED) on the web site of the National Map Seamless Server operated by
the U.S.\@ Geological Survey, EROS Data Center, Sioux Falls, South Dakota.
The particular dataset we used covers a region in the Appalachian mountains
on a $37 \times 23$ grid.
The elevation map is shown in Figure~\ref{fig:appal-datamap},
which also marks 20 randomly selected locations that provided synthetic
measurements.

We modeled the elevation, denoted by $\Y(\x)$, by the geostatistical
formulation described in Section~\ref{sec:parameterization}
with a linear trend function and geometric anisotropy.
The smoothness $\smoothness$ was fixed at 1.5.
This formulation has eight parameters:
$\trend_0$, $\trend_X$, $\trend_Y$, $\angle$,
$\scale_X$, $\scale_Y$, $\sd^2$, and $\nugget$.
As mentioned in Section~\ref{sec:algorithm},
the algorithm works with a transformed parameter vector:
$\theta = \bigl(\trend_0, \trend_X, \trend_Y,
\log \frac{\angle}{\frac{\pi}{2} - \angle},
\log \scale_X, \log \scale_Y, \log \sd^2,
\log \frac{\nugget}{1 - \nugget}\bigr)$.
It is straightforward to study the parameters
in their natural units by back-transforming samples
from the estimated posterior distribution of $\theta$.

The initial approximation $f^{(0)}(\theta)$ was taken
to be the product of eight independent and fairly diffuse
normal distributions,
one for each component of $\theta$.
This initial distribution was updated eight times
in the iterative algorithm.
During the iterations,
the approximate posterior, $f^{(k)}$,
converged to the true posterior, $\pi(\theta)\, p(\vy \given \theta)$
(up to a normalizing factor).
The convergence was examined via two measures
that indicate the ``closeness'' or ``distance'' of two
distributions.

The first measure is the entropy of importance sampling weights.
Consider the sample $\{\theta_i\}_{i=1}^n$ obtained in step~1
(see Section~\ref{sec:algor})
of the $k$-th iteration, which is a random sample
from the density $f^{(k-1)}(\theta)$.
The entropy of the importance weights
$\{w_i\}_{i=1}^n$, obtained in step~2 of the algorithm,
is the $\gamma$ defined in~(\ref{eq:flatten-by-entropy}).
The entropy can be used as an indicator of how close the
estimated distribution, $f^{(k-1)}(\theta)$,
is to the true posterior distribution.
An entropy value close to 1 indicates good approximation
\citep{West:1993:APD, Liu:1998:RCS}.

The second measure is the $L_1$ distance
between the estimated posterior,
say $f(\theta)$,
and the true posterior,
$c g(\theta) = c \pi(\theta) p(\vy \given \theta)$,
where $c$ is an unknown normalizing constant.
The $L_1$ distance is defined as
$L_1 = \int_{\Theta} \lvert f(\theta) - c g(\theta)\rvert \diff \theta$,
hence
\[
\begin{split}
L_1
&= \int_{\Theta}
    \biggl\lvert
        1 -
        \frac{g(\theta)}{f(\theta) \int_{\Theta} g(\theta) \diff\theta}
    \biggr\rvert
    f(\theta) \diff \theta
\\
&= \int_{\Theta}
    \biggl\lvert
        1 -
        \frac{\frac{g(\theta)}{f(\theta)}}
            {\int_{\Theta} \frac{g(\theta)}{f(\theta)} f(\theta) \diff\theta}
    \biggr\rvert
    f(\theta) \diff \theta
\\
&= E_{f(\theta)} \biggl\lvert
    1 -
    \frac{g(\theta)/f(\theta)}
        {E_{f(\theta)} \bigl[g(\theta)/f(\theta)\bigr]}
    \biggr\rvert
,
\end{split}
\]
where the two expectations are with respect to the distribution
$f(\theta)$.
Therefore,
a Monte Carlo estimate of this distance is
\[
d_{L_1}
= \frac{1}{n} \sum_{i=1}^n |1 - nw_i|
,
\]
making use of the fact that the sample mean of
$\frac{g(\theta)}{f(\theta)}$ is
$\frac{1}{n} \sum_{i=1}^n w_i$, i.e.\@ $\frac{1}{n}$.

The values of $\gamma$ and $d_{L_1}$ in each iteration
are listed in Table~\ref{tab:appal-converge},
which shows definite trends of increase in $\gamma$
and decrease in $d_{L_1}$.

\begin{table}
\caption{Convergence of the approximate posterior to the true posterior
    in iterations of Example~1,
    as indicated by functions of the importance sampling weights.}
\label{tab:appal-converge}
\centering
\begin{tabular}{l|cccccccccc}
\br
iteration ($k$) & 1 & 2 & 3 & 4 & 5 & 6 & 7 & 8\\ \hline
sample size ($n$) & 3000 & 2800 & 2620 & 2458 & 2312 & 2181 & 2063 & 1957\\
entropy ($\gamma$) & 0.12 & 0.35 & 0.37  & 0.57 & 0.80 & 0.88 & 0.91 & 0.92\\
$L_1$ distance ($d_{L_1}$) & 1.99 & 1.92  & 1.86  & 1.68 & 1.22 &
    1.05 & 0.88 & 0.84\\
\br
\end{tabular}
\end{table}

\begin{table*}
\caption{Empirical mean and standard deviation of 1000 samples
    of the parameters (back-transformed to their natural units)
    drawn from the final posterior approximation in Example~1.}
\label{tab:appal-mean}
\centering
\begin{tabular}{l|ccccccccc}
\br
parameter  & $\trend_0$ & $\trend_X$ & $\trend_Y$
    & $\scale_X$ & $\scale_Y$ & $\alpha$
    & $\sd^2$ & $\nugget$ \\ \hline
sample mean & 941 & -9.01 & 0.81
    & 36.1 & 2.90 & 0.32
    & 4.95e04 & 0.06 \\
sample s.d. & 168 & 4.67 & 11.6
    & 30.7 & 2.15 & 0.08
    & 7.12e04 & 0.09\\
\br
\end{tabular}
\end{table*}

\begin{figure}
\centering
\includegraphics[scale=.8]{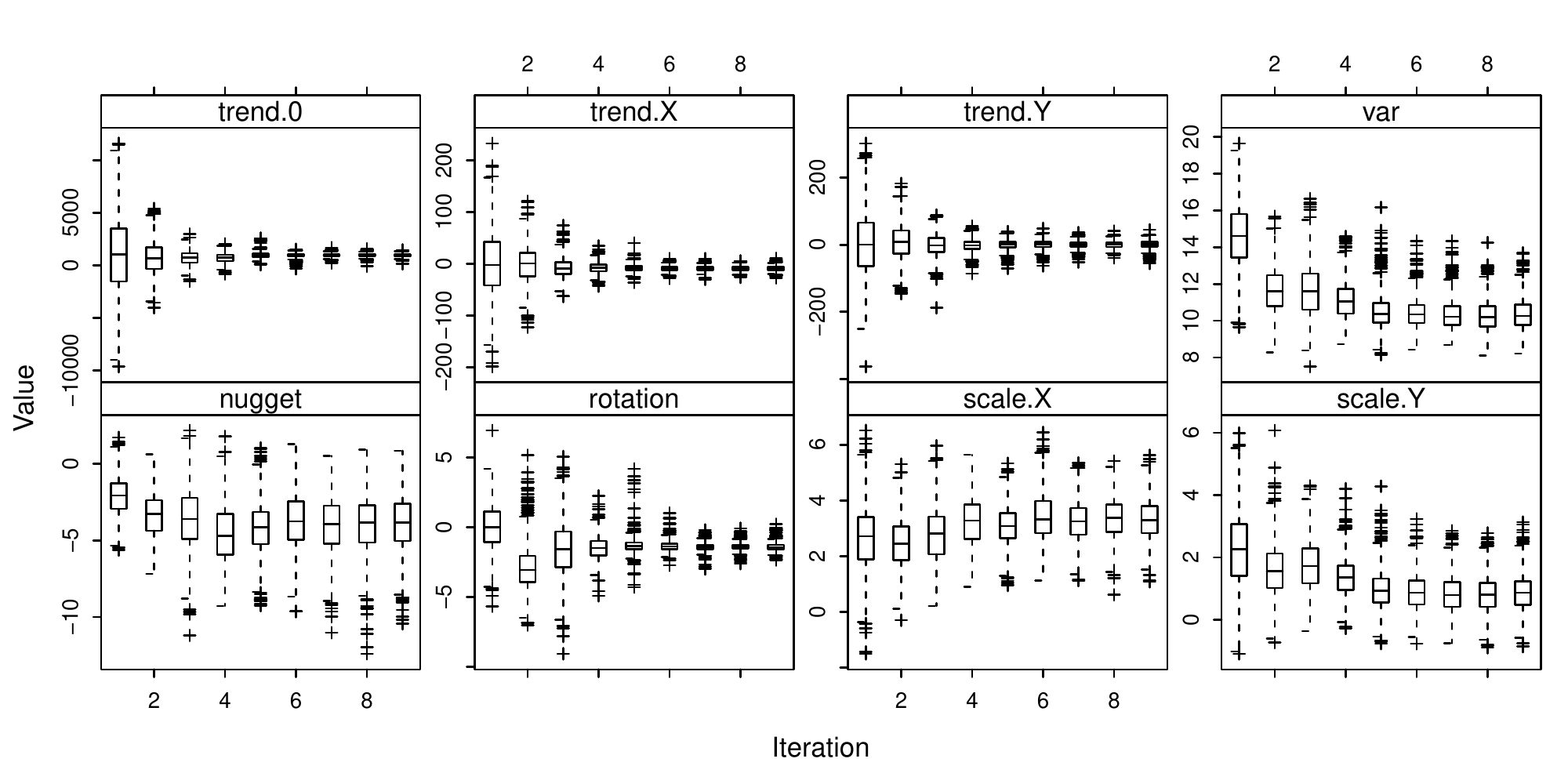}
\caption{Marginal distributions (by boxplots)
    of the model parameters as revealed by
    1000 samples drawn from each of the 8 iterative approximations
    to the posterior distribution.
    The parameters are in transformed units,
    hence taking values on $(-\infty, \infty)$.
    See Section~\ref{sec:example1}.}
\label{fig:appal-geost-marginal}
\end{figure}

In total we obtained nine approximate posterior distributions,
including the initial approximation and the eight subsequent updates.
From each of these approximations
we drew 1000 samples of the parameter vector $\theta$.
The marginal distribution of each component of $\theta$
is plotted in Figure~\ref{fig:appal-geost-marginal}.
It can be seen that the posterior approximations stabilized
after six or seven iterations,
consistent with the quantitative indicators in
Table~\ref{tab:appal-converge}.
The stabilized approximate distributions of the
scale, variance, and nugget parameters are more outspread
than those of the trend and rotation parameters.
This is related to the interactions between the former group of
parameters, which cause identifiability
difficulties \citep[see][]{Zhang:2004:IEA}.

To provide some idea of the posterior mean and uncertainty
of the individual parameter components,
Table~\ref{tab:appal-mean} lists
the empirical means and standard deviations of the eight parameter
components calculated using the 1000 samples from
$f^{(8)}(\theta)$, the final approximation.
The empirical posterior mean of the rotation, $\angle$,
is 0.32, i.e.\@ 18 degrees.
The empirical posterior mean of the scale along the rotated horizontal
axis ($\scale_X$) is 36.1, in contrast to 2.90 along the
vertical axis.
The angle and scales are in keeping with what we observe
in the synthetic true field.
We note that,
with the 20 irregularly located measurements in this example,
these anisotropy parameters would be difficult to
estimate by methods based on curve-fitting for empirical variograms
\citep[see, e.g.][]{Paleologos:2011:SAF}.

\begin{figure}
\centering
\includegraphics[scale=.9]{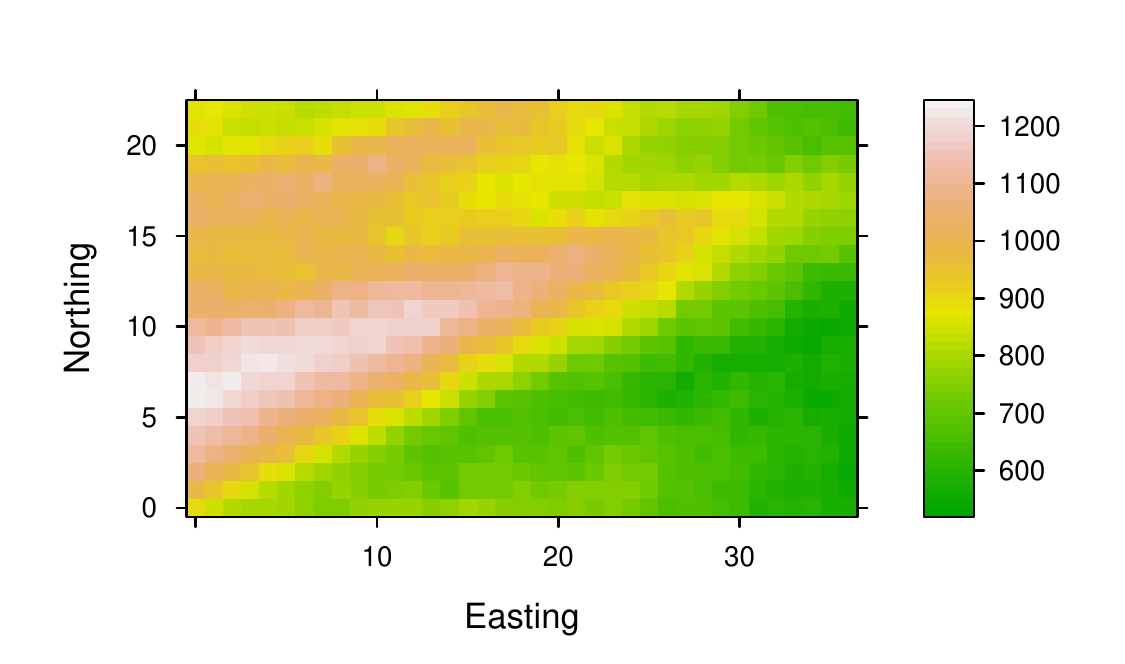}
\caption{Point-wise median of 100 simulations
    conditional on model parameters drawn from
    the final approximation to the posterior distribution.
    See Section~\ref{sec:example1}.}
\label{fig:appal-medianmap}
\end{figure}

\begin{figure}
\centering
\includegraphics[scale=.9]{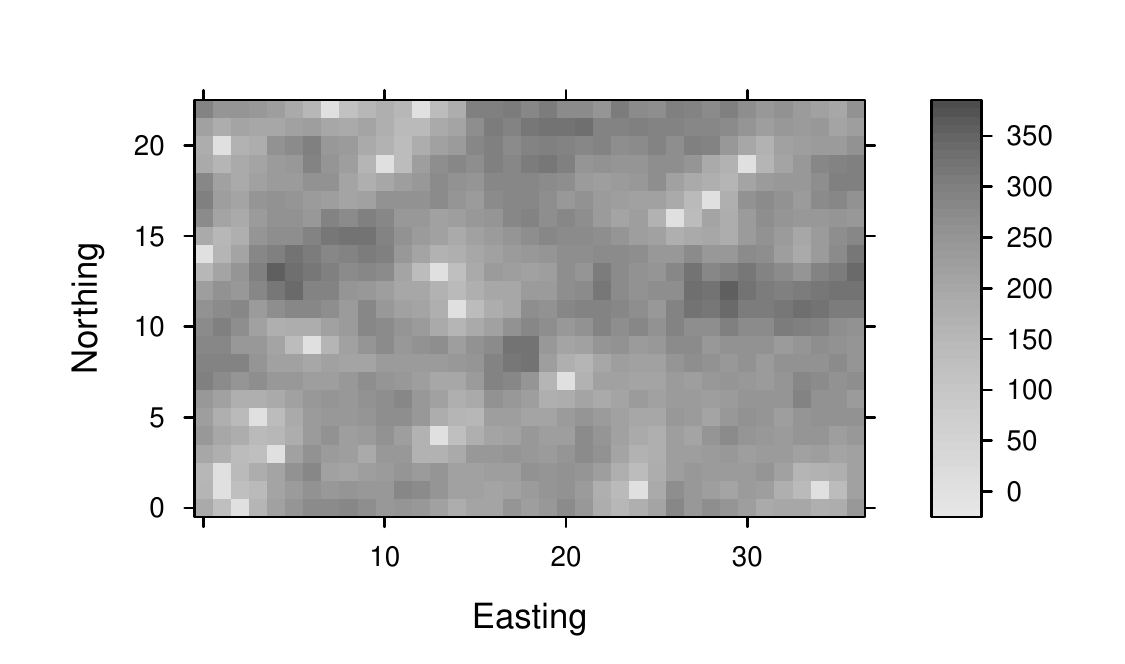}
\caption{Point-wise standard deviation of 100 simulations
    conditional on model parameters drawn from
    the final approximation to the posterior
    distribution. Note that variation at the locations of the 20
    measurements is 0.
    See Section~\ref{sec:example1}.}
\label{fig:appal-uncertaintymap}
\end{figure}

In geostatistical analysis,
one of the usual goals is to provide an interpolated map of the spatial field
accompanied by a measure of uncertainty.
We conducted 100 conditional simulations based on the final
approximation to the posterior distribution of the parameter.
The point-wise median of the simulations is shown
in Figure~\ref{fig:appal-medianmap}.
A comparison of Figure~\ref{fig:appal-medianmap} with
Figure~\ref{fig:appal-datamap} confirms that the model and algorithm
have captured main features of the true spatial field.
The point-wise standard deviation of the simulations
is shown in Figure~\ref{fig:appal-uncertaintymap}.
The level of uncertainty in the simulated fields is largely uniform
throughout the model domain, because the locations of the observations are
reasonably balanced in the model domain.

\subsection{Example 2}
\label{sec:example2}

\begin{figure}
\centering
\includegraphics[scale=.9]{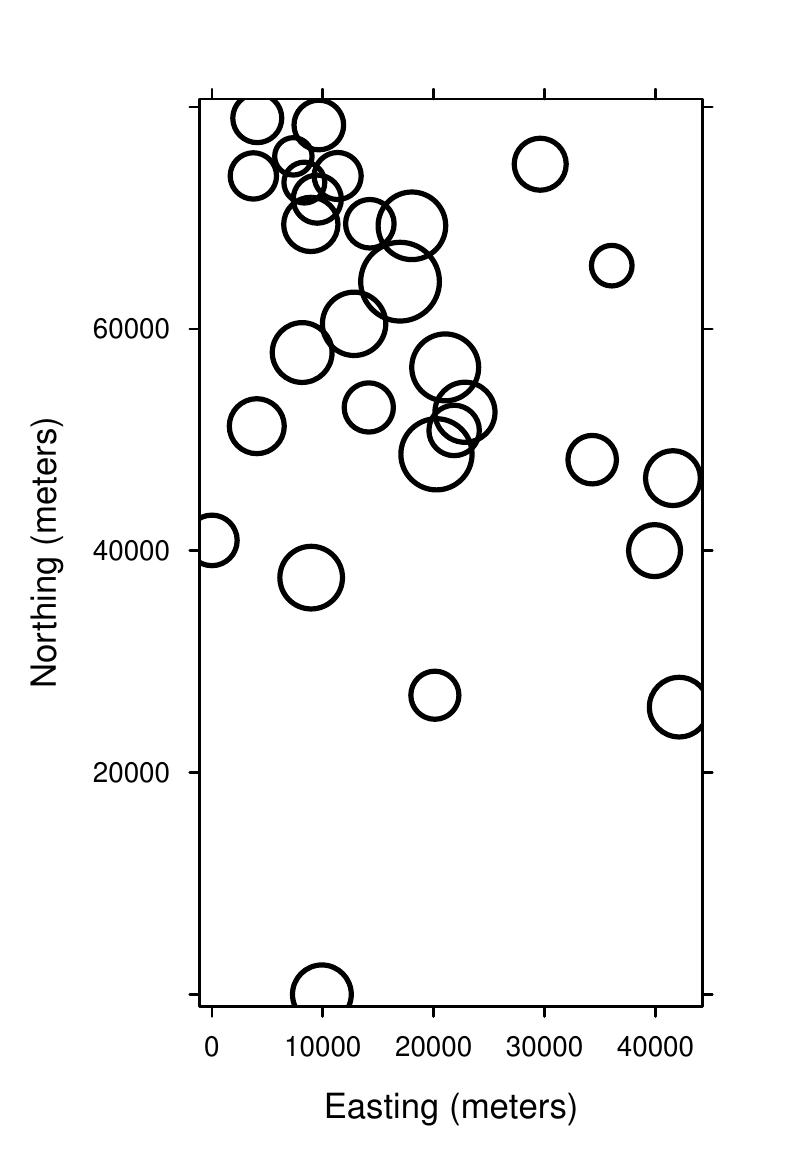}
\caption{Locations of 29 soil moisture measurements
    ($\unit{{cm}^3/{cm}^3}$) for the example in
    Section~\ref{sec:example2}.
    Areas of the circles are proportional to the moisture values.}
\label{fig:soil-datamap}
\end{figure}

In the second example,
we used a hydrologic dataset provided by
the Southeast Watershed Research Laboratory (SEWRL)
of the U.S.\@ Department of Agriculture.
The dataset contains long-term records of a number of hydrologic variables
for the Little River Experimental Watershed in south-central Georgia,
United States.
This research program as well as its data products are described in
\citet{Bosch:2007:LRE}.
For illustration,
we focused on a time snapshot
(specifically, at 18:00 on 1 August 2007)
of soil moisture represented
by measurements at 29 irregularly-located gauges.
The measurements are shown in
Figure~\ref{fig:soil-datamap}.
Details about the regional geography and the moisture data
can be found in
\citet{Bosch:2007:LRE} and
\citet{Bosch:2007:PSM}, respectively.

Because soil moisture is a percentage,
we took its logit transform as the spatial variable $Y$, that is,
$Y = \log \frac{\text{moisture}}{0.5 - \text{moisture}}$.
(The upper bound was taken to be 0.5 because it was noticed that
the largest observed value is 0.34.)
Such a transformation is needed because $Y$ must be defined on
$(-\infty,\infty)$ in order to be modeled as a Gaussian variable.
This treatment is in line with generalized linear models
in model-based geostatistics \citep{Diggle:2007:MBG};
but see \citet{Michalak:2005:MIN} for an alternative approach.

With the parameterization described in Section~\ref{sec:parameterization},
we took the trend model $\transpose{\mu(x)}\trend$
to be a constant $\trend$,
fixed the smoothness parameter $\smoothness$ at 0.5
(i.e.\@ used an exponential correlation function),
and did not consider anisotropy.
This left us with four parameters:
$\trend$, $\sd^2$, $\scale$, and $\nugget$.
The algorithm worked with a transformed parameter vector:
$\theta = \bigl(\trend, \log \sd^2, \log \scale,
\log \frac{\nugget}{1 - \nugget}\bigr)$.

Construction of the prior and the initial approximate posterior
distributions followed procedures similar to those in the first example.
The initial distribution was updated five times.
The convergence of the approximate posterior distributions to the truth
were examined via the two measures
$\gamma$ and $d_{L_1}$ as listed in
Table~\ref{tab:soil-converge}.
The values listed
show definite trends of increase in $\gamma$
and decrease in $d_{L_1}$.
As $\gamma$ and $d_{L_1}$ approached their respective limits, 1 and 0,
their values began to ``level off''.

\begin{table}
\caption{Convergence of the approximate posterior to the true posterior
    in iterations of Example~2,
    as indicated by functions of the importance sampling weights.}
\label{tab:soil-converge}
\centering
\begin{tabular}{l|ccccccc}
\br
iteration ($k$) & 1 & 2 & 3 & 4 & 5 \\ \hline
sample size ($n$) & 2000 & 1880 & 1772 & 1675 & 1587 \\
entropy ($\gamma$) & 0.47 & 0.92 & 0.98  & 0.98 & 0.99 \\
$L_1$ distance ($d_{L_1}$) & 1.82 & 0.87  & 0.40  & 0.36 & 0.31 \\
\br
\end{tabular}
\end{table}

\begin{table}
\caption{Empirical mean and standard deviation of 1000 samples
    of the parameters (back-transformed to their natural units)
    drawn from the final posterior approximation in Example~2.}
\label{tab:soil-mean}
\centering
\begin{tabular}{l|ccccccc}
\br
parameter  & $\trend$ & $\scale$ & $\sd^2$ & $\nugget$ \\ \hline
sample mean & -0.78 & 30897 & 0.52 & 0.21\\
sample s.d. & 0.39 & 56140 & 0.41 & 0.16\\
\br
\end{tabular}
\end{table}

Table~\ref{tab:soil-mean} lists
the empirical means and standard deviations of the four parameter
components obtained using 1000 samples from the final posterior
approximation.

\begin{figure}
\centering
\includegraphics[scale=.9]{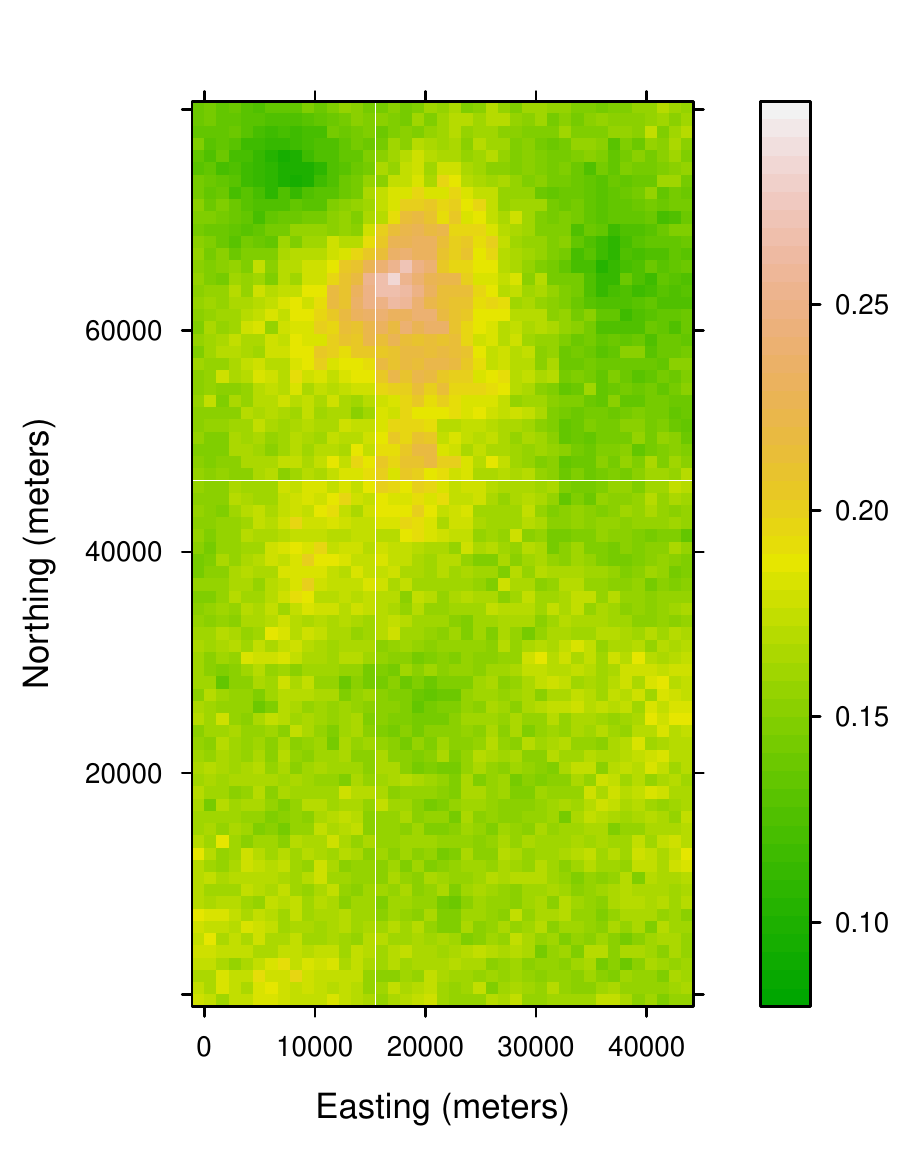}
\caption{Point-wise median of 100 simulations
    conditional on model parameters drawn from
    the final approximation to the posterior distribution.
    See Section~\ref{sec:example2}.}
\label{fig:soil-medianmap}
\end{figure}

A major purpose of a geostatistical analysis
like the current one is to generate an ensemble of soil moisture maps
to be used in subsequent studies that require
the value of soil moisture in the entire domain.
We divided the model domain into
41 (east-west) by 74 (north-south) grids,
each of size $1106\unit{m} \times 1106\unit{m}$,
and conducted 100 conditional simulations based on the approximate posterior
distribution obtained in the final iteration of the algorithm.
Each simulation was back-transformed to give a moisture map
with values in $(0,0.5)$,
in contrast to the variable $\Y \in (-\infty, \infty)$ in the model.
The point-wise median of the simulations is shown
in Figure~\ref{fig:soil-medianmap}, which confirms
a high-moisture area in the upper-central section of the model domain.
More scientific insights are expected if one
examines the simulated soil moisture maps
in the context of other hydrologic variables as well as the geography.

\section{Conclusion}
\label{sec:concl}

We have described a Bayesian geostatistical framework
and proposed an iterative algorithm for deriving
the posterior distribution of the parameter vector.
The contribution of this study is to provide a
general inference procedure that avoids some difficult elements
utilized in practice, including
(1) fitting a variogram curve;
(2) imposing bounds on an unbounded parameter;
(3) discretizing a continuous parameter.
Moreover,
the procedure can be applied to other model formulations
as long as the model parameters, or transforms thereof,
are defined on $(-\infty, \infty)$.
Common transformations that achieve this goal include
the logarithmic and logistic transformations,
as exemplified in Section~\ref{sec:examples}.

The algorithm centers on normal kernel density estimation.
Particular efforts are made to determine the localization and bandwidth
parameters in a systematic fashion.
Difficulties caused by highly skewed importance sampling weights
are alleviated by ``flattening'' the weights.

The method was demonstrated by two examples using synthetic and historical data.
In both examples,
we examined convergence of the approximate posterior distributions,
as well as features of the marginal posterior distributions.
The estimated posterior distributions served as a basis for
conditional simulations of the spatial field.

\bigskip

\noindent
\textbf{Acknowledgement:}\ \ 
{\small%
The author's Senior Visiting Scholarship at Tsinghua University was funded by
the Excellent State Key Lab Fund no.\@ 50823005,
National Natural Science Foundation of China,
and
the R\&D Special Fund for Public Welfare Industry no.\@ 201001080,
Chinese Ministry of Water Resources.
}


\newpage

\twocolumn

\end{document}